\documentclass[aps,preprint]{revtex4}%
\usepackage{amsfonts}
\usepackage{amsmath}
\usepackage{amssymb}
\usepackage{graphicx}%
\begin{document}
\preprint{ }
\title[ ]{Improved treatment of fermion-boson vertices and Bethe-Salpeter equations in
non-local extensions of dynamical mean field theory }
\author{A. Katanin}
\affiliation{Moscow Institute of Physics and Technology, Institutsky lane 9, Dolgoprudny, 141700, Moscow region, Russia\\
M. N. Mikheev Institute of Metal Physics, Kovalevskaya str. 18, 620990, Ekaterinburg, Russia}

\begin{abstract}
We reconsider the procedure of calculation of fermion-boson vertices and
numerical solution of Bethe-Salpeter equations, used in non-local extensions
of dynamical mean-field theory. Because of the frequency dependence of
vertices, finite frequency box for matrix inversions is typically used, which
often requires some treatment of asymptotic behavior of vertices. Recently
[Phys. Rev. B {\bf 83}, 085102 (2011); {\bf 97}, 235140 (2018)] it was proposed to split the considered
frequency box into smaller and larger one; in the smaller frequency box the
numerically exact vertices are used, while beyond this box asymptotics of
vertices are applied. Yet, this method requires numerical treatment of vertex
asymptotics (including corresponding matrix manipulations) in the larger
frequency box and/or knowing fermion-boson vertices, which may be not convenient for numerical calculations. In the
present paper we derive the formulae which treat analytically contribution of
vertices beyond chosen frequency box, such that only numerical operations
with vertices in the chosen small frequency box are required. The method is tested on
the Hubbard model and can be used in a broad range of applications of
non-local extensions of dynamical mean-field theory.

\end{abstract}
\maketitle

\section{Introduction}

Strongly-correlated systems show fascinating physical properties, like
coexistence of magnetic and charge correlations\cite{High-Tc1},
high-temperature superconductivity\cite{High-Tc1,High-Tc2,High-Tc}, Hund metal
behavior\cite{Hund}, etc. Local correlations, which appear due to the
(non-)local interactions in strongly-correlated systems are well described by
the (extended) dynamical mean-field theory ((E)DMFT)
\cite{DMFT,DMFT2,EDMFT,EDMFT_Si}. At the same time, this theory is not
sufficient to describe the non-local correlations, which play crucial role in
many phenomena in strongly-correlated systems, in particular in quantum and
classical phase transitions, superconductivity, etc. The non-local extensions
of dynamical mean field theory, such as dynamic cluster approximation
and cellular mean-field theory (see for a review Ref. \cite{RevCluster}) meet
difficulties when treat low temperatures and large cluster sizes. Recent
progress in diagrammatic extensions of (E)DMFT \cite{Review}, namely ladder
\cite{DGA1a,DGA1b,DGA1c,DGA1d,DGA2,abinitioDGA} and parquet \cite{ParquetDGA}
dynamic vertex approximation (D$\Gamma$A), dual fermion (DF) approach
\cite{DF1,DF2,DF3,DF4}, dual boson (DB) approach \cite{DB1,DB2,DB3,DB4},
TRILEX \cite{TRILEX}, DMF$^{2}$RG approach\cite{DMF2RG,DMF2RG3} and
(E)DMFT+2PI-fRG method \cite{MyEDMFT2PI} allowed to treat non-local
correlations on a non-perturbative basis.

Key ingredient of many of these methods is the relation between given
two-particle irreducible vertices (which are often assumed to be local) and
the two-particle reducible vertices, expressed by the corresponding
Bethe-Salpeter equations, as well as calculation of the fermion-boson vertices \cite{DGA1c,DB4,TRILEX,MyEDMFT2PI,FermionBoson}. Due to using finite frequency box, the corresponding
treatment is, however, often approximate, and to get reasonable results large
frequency box is required, which makes numerical calculation of vertices
within this frequency box difficult. Recently it was proposed \cite{Kunes,Toschi} to
split the frequency box into ``small" one where the numerically exact vertices
are used, and larger one, where vertex asymptotics are used. The proposed
approach requires however numerical treatment of vertices in the large frequency box (although
with their asymptotic values) and/or knowing fermion-boson vertices, which make it not very convenient for applications.
In the present paper we propose a way of analytical treatment of vertex
asymptotics, such that only numerical calculations within small frequency box
are required.

The plan of the paper is the following. In Sect. II we introduce the model. In
Sect. III we consider procedure of calculation of fermion-boson vertices and
susceptibilities using the interaction vertex obtained in a given frequency
box. In Sect. IV we discuss the solution of the Bethe-Salpeter equation. In
Sect. V we present a numerical example of application of the obtained formulae
to the standard Hubbard model. In Sect. VI we present Conclusions.

\section{The model and asymptotics of vertices}

We consider an extended Hubbard model described by an action%
\begin{equation}
\mathcal{S}=-\sum\limits_{k,\sigma}c_{k\sigma}^{+}G_{0k}^{-1}c_{k\sigma}%
+U\sum\limits_{q}n_{q\uparrow}n_{-q,\downarrow}+\frac{1}{2}\sum\limits_{q}%
V_{q}^{c}n_{q}n_{-q},\label{S_L}%
\end{equation}
where $G_{0k}$ and $V_{q}^{c}$ are some (arbitrary) single-particle Green
function and the two-particle vertex, $c_{k\sigma}^{+},c_{k\sigma}$ are
Grassmann variables, $\sigma=\uparrow,\downarrow,$ $n_{q}=\sum
\nolimits_{\sigma}n_{q\sigma}=\sum\nolimits_{k,\sigma}c_{k\sigma}^{+}%
c_{k+q,\sigma},$ and we use the momentum-frequency variables $k=(\mathbf{k,}%
i\nu_{n}),$ $q=(\mathbf{q},i\omega_{n}),$ where $i\nu_{n}$ and $i\omega_{n}$ are
fermionic- and bosonic Matsubara frequencies. The action (\ref{S_L}) can
describe both, the (E)DMFT solution of the Hubbard model (in which case
$G_{0k}$ and $V_{q}^{c}$ are only frequency dependent), as well as more
general case of the non-local theory, for which $G_{0k}$ and/or $V_{q}^{c}$
acquire some momentum dependence.

Let us denote the full two-particle vertex in charge (c) and spin (s) channels (which
we consider below for definiteness), corresponding to the action (\ref{S_L})
by $\mathcal{F}_{\nu\nu^{\prime}q}^{c(s)}$ , where $\nu,\nu^{\prime}$ are the
incoming- and outgoing fermionic Matsubara frequencies, and $q$ is the momentum-frequency transfer. We assume for simplicity that
the vertex depends only on one of the momenta (i.e. the momentum transfer
$\mathbf{q}$), as it happens in the ladder versions of D$\Gamma$A
\cite{DGA1a,DGA1b,DGA1c,DGA1d,abinitioDGA}, DF \cite{DF3,DF4}, DB
\cite{DB2,DB3}, and (E)DMFT+2PI-fRG \cite{MyEDMFT2PI} approaches; more general
case can be treat in a similar way. The vertex $\mathcal{F}_{\nu\nu^{\prime}%
q}^{c(s)}$ is related to the two-particle irreducible vertex $\Phi_{\nu
\nu^{\prime}q}^{c(s)}$ by the Bethe-Salpeter equation%
\begin{equation}
\mathcal{F}_{\nu\nu^{\prime}q}^{c(s)}=\left[  (\Phi_{\nu\nu^{\prime}q}%
^{c(s)})^{-1}-\delta_{\nu\nu^{\prime}}\chi_{\nu q}^{0}\right]  _{\nu
\nu^{\prime}}^{-1},\label{BS1}%
\end{equation}
where $\chi_{\nu q}^{0}=-\sum\nolimits_{\mathbf{k}}G_{k}G_{k+q}$,
$G_{k}^{-1}=G_{0k}^{-1}-\Sigma_{k}$ is the full Green function and $\Sigma
_{k}$ is the electronic self-energy (for DMFT and ladder D$\Gamma$A the latter
depends on the fermionic frequency $\nu$ only). The vertex $\Phi_{\nu
\nu^{\prime}q}^{c(s)}$ for the considering cases of (E)DMFT and its non-local ladder extensions has at $\nu\rightarrow\infty$ or $\nu^{\prime
}\rightarrow\infty$ the asymptotic form \cite{Toschi}
\begin{equation}
\Phi_{\nu\nu^{\prime}q}^{c(s)}\rightarrow U_{q}^{c(s)}+\overline{\Phi}_{\nu
\nu^{\prime}\omega}^{c(s)},\label{Vertex_ir_as}%
\end{equation}
where $U_{q}^{c}=-(U+2V_{q}^{c}),$ $U_{q}^{s}=U,$ and  $\overline{\Phi}_{\nu
\nu^{\prime}\omega}^{c(s)}$ is given by
\begin{align}
\overline{\Phi}_{\nu\nu^{\prime}\omega}^{s} &  =-U^{2}[\chi_{c}(\nu
-\nu^{\prime})-\chi_{s}(\nu-\nu^{\prime})]/2-U^{2}\chi_{pp}(\nu+\nu^{\prime
}+\omega)+v^{c}(\nu-\nu^{\prime}),\label{Phi_bar}\\
\overline{\Phi}_{\nu\nu^{\prime}\omega}^{c} &  =-U^{2}[\chi_{c}(\nu-\nu
^{\prime})+3\chi_{s}(\nu-\nu^{\prime})]/2+U^{2}\chi_{pp}(\nu+\nu^{\prime
}+\omega)+v^{c}(\nu-\nu^{\prime}),\nonumber
\end{align}
$\chi_{c,s,pp}(\omega)$ are the charge-, spin-, and
particle-particle susceptibilities, accounting for the contribution of the respective bubbles in the transverse channel (these contributions are
assumed to be local in the considering ladder approximation), $v^{c}(\omega)$
is the local retarded Coulomb interaction, corresponding to the original non-local interaction $V_q^c$, and obtained, e.g., in EDMFT \cite{EDMFT,EDMFT_Si} (non-local corrections to this interaction w.r.t. $\nu$ and $\nu'$ are neglected in the considering case of EDMFT and its non-local ladder extensions). 
Note that $\overline{\Phi}_{\nu
\nu^{\prime}\omega}^{c(s)}$ can be calculated for arbitrary large $\nu,\nu^{\prime}$
since $\chi_{c,s,pp}(\omega)$ and $v^{c}(\omega)$ decay as $1/\omega^2$
(outside the bosonic frequency box they can be therefore approximated by zero or replaced by the respective asymptotic behavior).

The corresponding asymptotics of the reducible vertices $\mathcal{F}_{\nu
\nu^{\prime}q}^{c(s)}$ at large frequency $\nu$ or $\nu^{\prime}$ fulfill 
\begin{equation}
\mathcal{F}_{\nu\nu^{\prime}q}^{c(s)}\simeq\left\{
\begin{array}
[c]{cc}%
U_{q}^{c(s)}\Gamma_{\nu^{\prime}q}^{c(s)}+\overline{\Phi}_{\nu\nu^{\prime
}\omega}^{c(s)}+\sum\limits_{\nu^{\prime\prime}}\overline{\Phi}_{\nu
\nu^{\prime\prime}\omega}^{c(s)}\chi_{\nu^{\prime\prime}q}^{0}\mathcal{F}%
_{\nu^{\prime\prime}\nu^{\prime}q}^{c(s)}, & \nu\rightarrow\infty,\\
U_{q}^{c(s)}\Gamma_{\nu q}^{c(s)}+\overline{\Phi}_{\nu\nu^{\prime}\omega
}^{c(s)}+\sum\limits_{\nu^{\prime\prime}}\mathcal{F}_{\nu\nu^{\prime\prime}%
q}^{c(s)}\chi_{\nu^{\prime\prime}q}^{0}\overline{\Phi}_{\nu^{\prime\prime}%
\nu^{\prime}\omega}^{c(s)}, & \ \nu^{\prime}\rightarrow\infty,
\end{array}
\right.  \ \label{Vertex_as}%
\end{equation}
where the three-leg (fermion-boson) vertex $\Gamma_{\nu q}^{c(s)}$ is defined by%
\begin{equation}
\Gamma_{\nu q}^{c(s)}=1+\sum\limits_{\nu^{\prime}}\mathcal{F}_{\nu\nu^{\prime
}q}^{c(s)}\chi_{\nu^{\prime}q}^{0},\label{Lambda}%
\end{equation}
here and below we assume factor of
temperature $T$ for every frequency summation. Note that for completeness we account for the last terms in the right-hand
sides of Eqs. (\ref{Vertex_as}), which were omit in Ref. \cite{Toschi}. We
note also that our definition of vertices $\mathcal{F}_{\nu\nu^{\prime}%
q}^{c(s)}$ and ${\Phi}_{\nu\nu^{\prime}q}^{c(s)}$ has opposite sign in
comparison to that used in Ref. \cite{Toschi}, and the vertex $\Gamma_{\nu
q}^{c(s)}$ corresponds to the vertices $1\pm\lambda_{\nu q}^{c(s)}$ of that paper.

\section{Three-leg vertices and susceptibilities}

Our first task is to obtain a closed expression for $\Gamma_{\nu q}^{c(s)}$
containing only summation of vertices (except their asymptotic parts $\overline \Phi^{c(s)}_{\nu\nu'\omega}$) within a given frequency box $\nu^{\prime}\in B.$
For that we split a summation in Eq. (\ref{Lambda}) into $\nu^{\prime}\in B$
and $\nu^{\prime}\notin B$ and use the asymptotic form of Eq. (\ref{Vertex_as}%
). Since $\overline{\Phi}_{\nu^{\prime\prime}\nu^{\prime}\omega}^{c(s)}\propto 1/\nu^2$
for large $\nu=||\nu^{\prime\prime}|-|\nu^{\prime}||$ (see Eqs.
(\ref{Phi_bar})), it is sufficient to approximate $\mathcal{F}%
_{\nu\nu^{\prime\prime}q}^{c(s)}\simeq U_{q}^{c(s)}\Gamma_{\nu q}^{c(s)}$ in
the right-hand side of Eq. (\ref{Vertex_as}) to the accuracy $O(1/\nu_{\rm max}^3)$, where $\nu_{\rm max}$ is the size of the frequency box.  
Substituting this into Eq. (\ref{Lambda}) and splitting also the summation over $\nu''$ in Eq. (\ref{Vertex_as}) into one inside and outside the frequency box, we obtain%
\begin{equation}
\Gamma_{\nu q}^{c(s)}=Z_{\nu q}^{c(s)}+\sum\limits_{\nu^{\prime}\in B}Z_{\nu' q}^{c(s)}\mathcal{F}_{\nu
\nu^{\prime}q}^{c(s)}\chi_{\nu^{\prime}q}^{0}+U_{q}^{c(s)}\Gamma_{\nu
q}^{c(s)}X_{q}^{c(s)},
\end{equation}
where $X_{q}^{c(s)}=\sum\nolimits_{\nu^{\prime}\notin B}\chi_{\nu^{\prime}q}^{0}%
+\sum\nolimits_{\nu^{\prime},\nu^{\prime\prime}\notin B}\chi_{\nu
^{\prime\prime}q}^{0}\overline{\Phi}_{\nu^{\prime\prime}\nu^{\prime}\omega
}^{c(s)}\chi_{\nu^\prime q}^{0},$ $Z_{\nu q}^{c(s)}=1+\sum\nolimits_{\nu^{\prime}\notin
B}\overline{\Phi}_{\nu\nu^{\prime}\omega}^{c(s)}\chi_{\nu^{\prime}q}^{0}$. From this
equation we find%
\begin{equation}
\Gamma_{\nu q}^{c(s)}=\frac{Z_{\nu q}^{c(s)}+\sum\limits_{\nu^{\prime}\in B}Z_{\nu' q}^{c(s)}\mathcal{F}%
_{\nu\nu^{\prime}q}^{c(s)}\chi_{\nu^{\prime}q}^{0}}{1-U_{q}%
^{c(s)}X_{q}^{c(s)}}.\label{Lambda_box}%
\end{equation}
The expression (\ref{Lambda_box}) gives a possibility to calculate
$\Gamma_{\nu q}^{c(s)}$ using summations of vertices (except their asymptotic parts) within the selected frequency box only. Using that $\chi^0_{\nu q}\propto 1/\nu^2$ for large $\nu$ and $\overline{\Phi}_{\nu\nu^{\prime}\omega}^{c(s)}\propto (|\nu|-|\nu^{\prime}|)^{-2}$ for large  $||\nu|-|\nu^{\prime}||$, we find that the first and second term in $X_{q}^{c(s)}$ are of the order $1/\nu_{\rm max}$ and $1/\nu_{\rm max}^2$, respectively,
and the difference $Z_{\nu q}^{c(s)}-1\propto 1/\nu_{\rm max}^3$, such that the second term in $X_q^{c(s)}$ and the difference $Z_{\nu q}^{c(s)}-1$ are expected to give only
very small contribution for large $\nu_{\rm max}$ (the smallness of these contributions is also verified to hold numerically for the DMFT solution of single band Hubbard model with some exemplary parameters, e.g. fillings close to half-filling, in Sec. V). 

Using the obtained fermion-boson vertex, we can similarly find the
non-local susceptibilities (which in general should not be confused with the
local susceptibilities $\chi^{c(s)}(\omega)$ entering Eq. (\ref{Phi_bar})) by
splitting again the summation inside and outside the frequency box:%
\begin{equation}
\chi_{q}^{c(s)}=\sum\limits_{\nu}\Gamma_{\nu q}^{c(s)}\chi_{\nu q}^{0}%
=\sum\limits_{\nu\in B}\Gamma_{\nu q}^{c(s)}\chi_{\nu q}^{0}+\sum
\limits_{\nu\notin B}\Gamma_{\nu,q}^{c(s)}\chi_{\nu q}^{0}.\label{chi}%
\end{equation}
Performing similar decomposition for $\Gamma_{\nu\notin B,q}^{c(s)}$ in Eq. (\ref{Lambda}) and using again Eq. (\ref{Vertex_as}) we find%
\begin{align}
\Gamma_{\nu\notin B,q}^{c(s)} &  \simeq Z_{\nu q}^{c(s)}
\left[  1+U_{q}^{c(s)}\chi_{q}^{c(s)}\right] +\sum\limits_{\nu' \in B} \overline \Phi_{\nu\nu'\omega}^{c(s)}\chi^0_{\nu' q} \Gamma_{\nu' q} \nonumber\\
&  \overset{\nu\rightarrow\infty}{\rightarrow}1+U_{q}^{c(s)}\chi_{q}%
^{c(s)}.\label{GammaAs1}%
\end{align}
Combining Eq. (\ref{chi}) with the first line of Eq. (\ref{GammaAs1}) this  yields  %
\begin{equation}
\chi_{q}^{c(s)}=\frac{\sum\limits_{\nu\in B}Z_{\nu q}^{c(s)}\Gamma_{\nu q}^{c(s)}\chi_{\nu
q}^{0}+X_{q}^{c(s)}}{1-U_{q}^{c(s)}X_{q}^{c(s)}},\label{chi1}%
\end{equation}
which again uses the summation only in a given frequency box. From Eq.
(\ref{chi1}) we find%
\begin{equation}
1+U_{q}^{c(s)}\chi_{q}^{c(s)}=\frac{1+U_{q}^{c(s)}\sum\limits_{\nu\in B}Z_{\nu q}^{c(s)}%
\Gamma_{\nu q}^{c(s)}\chi_{\nu q}^{0}}{1-U_{q}^{c(s)}X_{q}^{c(s)}}.\label{chi2}
\end{equation}
We have verified that the result in the first line of Eq. (\ref{GammaAs1}) with account of Eq. (\ref{chi2}) is identical to the large $\nu$ limit of Eq. (\ref{Lambda_box}).

Let us also consider the \textquotedblleft reduced\textquotedblright%
\ fermion-boson vertex \cite{EdwHertz}
\begin{equation}
\gamma_{\nu q}^{c(s)}=\frac{\Gamma_{\nu q}^{c(s)}}{1+U_{q}^{c(s)}\chi
_{q}^{c(s)}}, \label{gamma0}
\end{equation}
which contains the sum of contributions from 2PI vertices with excluded
$U_{q}^{c(s)}$ interaction. This vertex is often used in D$\Gamma$A
\cite{DGA1c}, TRILEX \cite{TRILEX}, some versions of the DB approach
\cite{DB4}, (E)DMFT+2PI-fRG method \cite{MyEDMFT2PI}, etc. For this vertex we
obtain%
\begin{align}
\gamma_{\nu q}^{c(s)} &  =\frac{Z_{\nu q}^{c(s)}+\sum\limits_{\nu^{\prime}\in B}%
\mathcal{F}_{\nu\nu^{\prime}q}^{c(s)}\chi_{\nu^{\prime}q}^{0}Z_{\nu' q}^{c(s)}%
}{1+U_{q}^{c(s)}\sum\limits_{\nu\in B}Z_{\nu q}^{c(s)}\Gamma_{\nu q}^{c(s)}\chi_{\nu q}^{0}%
}\nonumber\\
&  =\frac{Z_{\nu q}^{c(s)}+\sum\limits_{\nu^{\prime}\in B}\mathcal{F}_{\nu\nu^{\prime}%
q}^{c(s)}\chi_{\nu^{\prime}q}^{0}Z_{\nu' q}^{c(s)}}{1+\widetilde{U}_{q}^{c(s)}%
\sum\limits_{\nu\in B}Z_{\nu q}^{c(s)}\left\{ Z_{\nu
q}^{c(s)}+\sum\limits_{\nu^{\prime}\in B}%
\mathcal{F}_{\nu\nu^{\prime}q}^{c(s)}\chi_{\nu^{\prime}q}^{0}Z_{\nu' q}^{c(s)}\right\}  \chi_{\nu q}^{0}},\label{gamma}%
\end{align}
where $\widetilde{U}_{q}^{c(s)}=U_{q}^{c(s)}/(1-U_{q}^{c(s)}X_{q}^{c(s)}).$ According
to the Eq. (\ref{GammaAs1}),
\begin{equation}
\gamma_{\nu\notin B,q}^{c(s)}\simeq Z_{\nu,q}^{c(s)}+\sum\limits_{\nu' \in B} \overline \Phi_{\nu\nu'\omega}^{c(s)}\chi^0_{\nu' q} \gamma^{c(s)}_{\nu' q}\overset{\nu\rightarrow
\infty}{\rightarrow}1.
\end{equation}
For the irreducible susceptibility $\phi_{q}^{c(s)},$ which is related to the
non-local susceptibility $\chi_{q}^{c(s)}$ by
\begin{equation}
\chi_{q}^{c(s)}=\frac{\phi_{q}^{c(s)}}{1-U_{q}^{c(s)}\phi_{q}^{c(s)}}, \label{locvsir}
\end{equation}
we find%
\begin{align}
\phi_{q}^{c(s)} &  =\frac{\chi_{q}^{c(s)}}{1+U_{q}^{c(s)}\chi_{q}^{c(s)}%
}\label{phi0}\\
&  =\frac{\sum\limits_{\nu\in B}Z_{\nu q}^{c(s)}\Gamma_{\nu q}^{c(s)}\chi_{\nu q}^{0}+X_{q}^{c(s)}%
}{1+U_{q}^{c(s)}\sum\limits_{\nu\in B}Z_{\nu q}^{c(s)}\Gamma_{\nu q}^{c(s)}\chi_{\nu q}^{0}%
}.\label{phi}%
\end{align}
It can be verified by direct algebraic transformations that the obtained
quantities fulfill the result for the irreducible susceptibility, which follows from the Eqs. (\ref{chi}), (\ref{gamma0}), and (\ref{phi0}), cf. Ref. \cite{EdwHertz},
\begin{equation}
\sum\limits_{\nu}\gamma_{\nu q}^{c(s)}\chi_{\nu q}^{0}=\sum\limits_{\nu\in
B}Z_{\nu q}^{c(s)}\gamma_{\nu q}^{c(s)}\chi_{\nu q}^{0}+X_{q}^{c(s)}=\phi_{q}^{c(s)}.\label{phi1}%
\end{equation}
For the following it is convenient to represent the vertex $\mathcal{F}%
_{\nu\nu^{\prime}q}^{c(s)}$ via Bethe-Salpeter equation, similar to
(\ref{BS1}),
\begin{equation}
\mathcal{F}_{\nu\nu^{\prime}q}^{c(s)}=\left\{  \left[  \Phi_{\nu\nu^{\prime}%
q}^{c(s),\mathrm{box}}\right]  ^{-1}-\delta_{\nu\nu^{\prime}}\chi_{\nu q}%
^{0}\right\}  _{\nu\nu^{\prime}}^{-1},\label{BS2}%
\end{equation}
but with the inversion performed for $\nu,\nu^{\prime}\in B$ only (which
provides the difference between $\Phi_{\nu\nu^{\prime}q}^{c(s),\mathrm{box}}$
and $\Phi_{\nu\nu^{\prime}q}^{c(s)}$, see Sect. IV). Using this equation and performing algebraic manipulations, similar to those described in Appendix
C of Ref. \cite{MyEDMFT2PI}, the result (\ref{gamma}) can be represented in a
simpler form%
\begin{equation}
\gamma_{\nu q}^{c(s)}=\sum\limits_{\nu^{\prime}\in B}\left[  1-\left(
\Phi_{\nu\nu^{\prime}q}^{c(s),\mathrm{box}}-Z_{\nu q}^{c(s)}\widetilde{U}_{q}^{c(s)}Z_{\nu' q}^{c(s)}\right)
\chi_{\nu^{\prime}q}^{0}\right]_{\nu \nu'}  ^{-1}Z_{\nu' q}^{c(s)},\label{gamma1}%
\end{equation}
where again the inversion is performed for $\nu,\nu^{\prime}\in B$. This
result allows us to obtain fermion-boson vertices $\gamma_{\nu q}^{c(s)}$ by
performing summation over frequencies within the chosen frequency box. The
size of the frequency box should be such that the asymptotic (\ref{Vertex_as})
is reached close to the boundary of the frequency box. We also note that different way of efficient
calculation of fermion-boson vertices and irreducible susceptibilities in a non-local theory from the known local ones was suggested in Ref. \cite{Krien}.

\section{Bethe-Salpeter equation}

Now we consider the solution of the Bethe-Salpeter equation (\ref{BS1}) which
we write in the form%
\begin{equation}
\mathcal{F}_{\nu\nu^{\prime}q}^{c(s)}=\Phi_{\nu\nu^{\prime}q}^{c(s)}%
+\sum\limits_{\nu^{\prime\prime}}\Phi_{\nu\nu^{\prime\prime}q}^{c(s)}\chi
_{\nu^{\prime\prime}q}^{0}\mathcal{F}_{\nu^{\prime\prime}\nu^{\prime}q}%
^{c(s)}.\label{BS}%
\end{equation}
Splitting again the summation to the one restricted to the frequency box and outside the
box and using the asymptotic forms (\ref{Vertex_ir_as}) and (\ref{Vertex_as}),
we find
\begin{align}
\mathcal{F}_{\nu\nu^{\prime}q}^{c(s)} &  =\Phi_{\nu\nu^{\prime}q}^{c(s)}%
+\sum\limits_{\nu^{\prime\prime}\in B}\Phi_{\nu\nu^{\prime\prime}q}^{c(s)}%
\chi_{\nu^{\prime\prime}q}^{0}\mathcal{F}_{\nu^{\prime\prime}\nu^{\prime}%
q}^{c(s)}\nonumber\\
&  +\sum\limits_{\nu^{\prime\prime}\notin B}\left[  U_{q}^{c(s)}%
+\overline{\Phi}_{\nu\nu^{\prime\prime}\omega}^{c(s)}\right]  \chi_{\nu
^{\prime\prime}q}^{0}\left[  U_{q}^{c(s)}\Gamma_{\nu^{\prime}q}^{c(s)}%
+\overline{\Phi}_{\nu^{\prime\prime}\nu^{\prime}\omega}^{c(s)}+\overline{\Phi}%
_{\nu^{\prime\prime}\widetilde{\nu}^{\prime\prime}\omega}^{c(s)}\chi_{\widetilde
{\nu}^{\prime\prime}q}^{0}\mathcal{F}_{\widetilde{\nu}^{\prime\prime}%
\nu^{\prime}q}^{c(s)}\right]  .
\end{align}
From this equation we can express $\Phi_{\nu\nu^{\prime}q}^{c(s)}:$
\begin{align}
\Phi_{\nu\nu^{\prime}q}^{c(s)} &  =\sum\limits_{\nu^{\prime\prime}\in
B}\left\{  \mathcal{F}_{\nu\nu^{\prime\prime}q}^{c(s)}-\sum\limits_{\widetilde
{\nu}^{\prime\prime}\notin B}\left[  U_{q}^{c(s)}+\overline{\Phi}%
_{\nu\widetilde{\nu}^{\prime\prime}\omega}^{c(s)}\right]  \chi_{\widetilde{\nu
}^{\prime\prime}q}^{0}\right.  \nonumber\\
&  \left.  \times\left[  U_{q}^{c(s)}\Gamma_{\nu^{\prime\prime}q}%
^{c(s)}+\overline{\Phi}_{\widetilde{\nu}^{\prime\prime}\widetilde{\nu}%
^{\prime\prime\prime}\omega}^{c(s)}\left(  \delta_{\widetilde{\nu}^{\prime
\prime\prime}\nu^{\prime\prime}}+\chi_{\widetilde{\nu}^{\prime\prime\prime}%
q}^{0}\mathcal{F}_{\widetilde{\nu}^{\prime\prime\prime}\nu^{\prime\prime}%
q}^{c(s)}\right)  \right]  \right\}  \left[  \delta_{\nu^{\prime\prime}%
\nu^{\prime}}+\chi_{\nu^{\prime\prime}q}^{0}\mathcal{F}_{\nu^{\prime\prime}%
\nu^{\prime}q}^{c(s)}\right]  _{\nu^{\prime\prime}\nu^{\prime}}^{-1}%
\nonumber\\
&  =\sum\limits_{\nu^{\prime\prime}\in B}\mathcal{F}_{\nu\nu^{\prime\prime}%
q}^{c(s)}\left[  \delta_{\nu^{\prime\prime}\nu^{\prime}}+\chi_{\nu
^{\prime\prime}q}^{0}\mathcal{F}_{\nu^{\prime\prime}\nu^{\prime}q}%
^{c(s)}\right]  _{\nu^{\prime\prime}\nu^{\prime}}^{-1}\nonumber\\
&  -\sum\limits_{\nu^{\prime\prime}\notin B}\left[  U_{q}^{c(s)}%
+\overline{\Phi}_{\nu\nu^{\prime\prime}\omega}^{c(s)}\right]  \chi_{\nu
^{\prime\prime}q}^{0}\left\{  Z_{\nu 
^{{\prime}{\prime}}q}^{c(s)}\widetilde{U}_{q}^{c(s)}Z_{\nu
^{\prime}q}^{c(s)}  +\overline{\Phi}_{\nu^{\prime\prime}\nu^{\prime}\omega}%
^{c(s)}\right\}  ,
\end{align}
where we have used the result for the fermion-boson vertex (\ref{Lambda_box}) and neglected the terms of higher order than $1/\nu_{\rm max}^3$.
Finally, using again the Bethe-Salpeter equation (\ref{BS2}) and performing algebraic transformations, we obtain
\begin{equation}
\Phi_{\nu\nu^{\prime}q}^{c(s)}=\Phi_{\nu\nu^{\prime}q}^{c(s),\mathrm{box}%
}+U_{q}^{c(s)}-Z_{\nu q}^{c(s)}\widetilde{U}_{q}^{c(s)}Z_{\nu' q}^{c(s)},\label{Phi}%
\end{equation}
where we have again neglected the terms 
of the order $o(1/\nu_{\rm max}^3)$. The result (\ref{Phi}) can be also derived from \textquotedblleft Method
2\textquotedblright\ of Ref. \cite{Toschi} which uses 
$\mathcal F$'s asymptotics (Eq. (19) of that paper) by applying Eqs. (\ref{Vertex_ir_as}), (\ref{Vertex_as}), and (\ref{Lambda_box}) above. The relation (\ref{Phi}) allows one to find the ``physical'' 2PI vertex from given vertex  
$\mathcal{F}_{\nu\nu^{\prime}q}^{c(s)}$ which is known inside
the frequency box ($\nu,\nu^{\prime}\in B$) by exploiting the equation
(\ref{BS2}) for the
vertex $\Phi_{\nu\nu^{\prime}q}^{c(s),\mathrm{box}}$. 
On the other hand, knowing the vertex
$\Phi_{\nu\nu^{\prime}q}^{c(s)}$ and proceeding the reverse way one can find
the corresponding vertex $\mathcal{F}_{\nu\nu^{\prime}q}^{c(s)}.$ In the
ladder approximation the vertex $\Phi_{\nu\nu^{\prime}q}^{c(s)}$ is assumed to
be local and the same for the local and non-local problems. This allows one to
find the relation between the respective vertices $\Phi_{\nu\nu^{\prime}%
\omega}^{c(s),\mathrm{box}}$ and $\Phi_{\nu\nu^{\prime}q}^{c(s),\mathrm{box}}$
of the local and non-local problem, which are different because of slight
difference of $X_q^{c(s)}$ and $Z_{\nu q}^{c(s)}$. 
The equation (\ref{Phi}) also provides
explanation of the result (\ref{gamma1}) for the fermion-boson vertex: since $\Phi^{c(s) \mathrm{box}}%
_{\nu\nu^{\prime}q}-Z_{\nu q}^{c(s)}\widetilde{U}^{c(s)}_{q}Z_{\nu' q}^{c(s)}=\Phi^{c(s)}_{\nu\nu^{\prime}%
q}-U^{c(s)}_{q}$, the obtained vertex $\gamma_{\nu q}$ in terms of the physical
vertex $\Phi^{c(s)}$ has a rather standard form (cf. Ref. \cite{DGA1b}), which is due to smallness of 
the difference $\Phi^{c(s)}_{\nu\nu^{\prime}q}-U^{c(s)}_{q}$ in the
limit $\nu\rightarrow\infty$ or $\nu^{\prime}\rightarrow\infty$; 
the factors $Z_{\nu q}^{c(s)}$ play the role of additional vertex corrections, accounting for the finite size of the frequency box.

In the approximation $Z_{\nu q}^{c(s)}\approx 1$ (which implies neglect of the contributions to the vertex of the order $1/\nu_{\rm max}^3$) the result of Eq. (\ref{Phi}) implies that the physical 2PI vertex and the 2PI vertex obtained in the
frequency box via Eq. (\ref{BS2}) differ by a q-dependent shift only. The quality of this approximation is verified for the Hubbard model near half filling in Sect. V. 

Obtained results allow to compare the accuracy of vertex calculation with and without the obtained corrections for the size of the frequency box. Without using the obtained corrections the main neglected contribution to the considered vertices arise from the terms, containing $X_{q}^{c(s)}\sim 1/\nu_{\mathrm{max}}$. Therefore, in this case the error of estimating two-particle irreducible and fermion-boson vertices also scales as $ 1/\nu_{\mathrm{max}}$. At the same time, accounting for the obtained corrections, the main source of the error of estimating of considered vertices is the deviation of irreducible vertices from asymptotic behavior  (\ref{Vertex_ir_as}) which is expected to scale as $1/{\rm max}(|\nu|,|\nu'|)^3$ and yield $O(1/\nu_{\mathrm{max}}^4)$ corrections to the vertex. Considering that the contribution of the terms of the order $1/\nu_{\rm max}^2$ (second term in $X_q^{c(s)}$) and $1/\nu_{\rm max}^3$ (i.e. $Z_{\nu q}^{c(s)}-1$) to the vertices is small (see also numerical verification in Sect. IV), higher order terms are expected to provide also small contribution, and therefore the suggested method provides fast convergence with increasing size of the box, as verified numerically in the next Section. 

\section{Numerical example}

\begin{figure}[t]
\center \includegraphics[width=0.95 \linewidth]{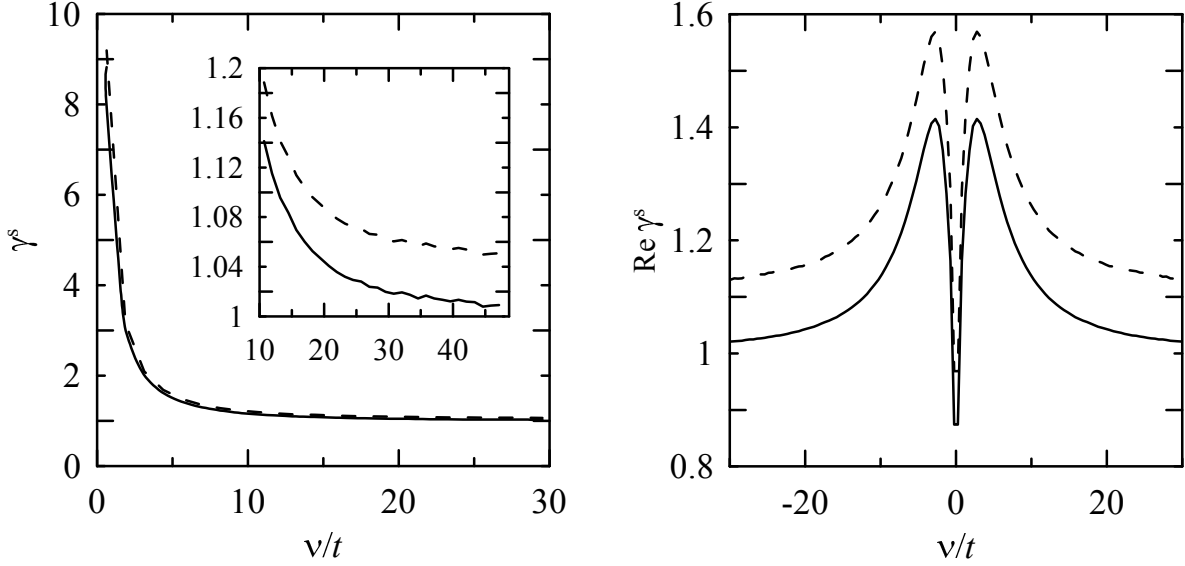} \endcenter
\caption{Real part of the fermion-boson vertex $\gamma^{s}_{\nu,0}$ in DMFT for
two-dimensional Hubbard model ($t^{\prime}=0.15t$, $U=10t$) for $T=0.2t$,
$n=1$ (left) and $T=0.08t$ $n=0.96$ (right). Dashed lines correspond to the
calculation, performed only within the chosen frequency box, while solid lines
show the result according to the Eq. (\ref{gamma}). The left plot is shown only for $\nu>0$ (the considered real part of the vertex is an even function of the frequency), the inset shows zoom of the asymptotic behavior at large frequencies.}%
\label{FigGamma}%
\end{figure}

As an example of the application of the developed approach we calculate the
spin vertex $\gamma^{s}_{\nu,0}$ in DMFT approach for the two-dimensional
Hubbard model with the dispersion $\epsilon_{\mathbf{k}}=-2t(\cos k_{x}+\cos
k_{y})+4t^{\prime}\cos k_{x} \cos k_{y}$. We choose the parameters $t^{\prime
}=0.15t$ and $U=10t$, which were suggested previously to describe physical
properties of high-$T_{c}$ compound La$_{2-x}$Sr$_{x}$CuO$_{4}$. For numerical
implementation of DMFT we use hybridization expansion continous-time QMC
method within iQIST package of Refs. \cite{iQIST,iQIST1}, choosing for the
frequency box $N_{f}=120$ fermionic Matsubara frequencies.

In the left part of Fig. \ref{FigGamma} we show the result of the calculation
of fermion-boson vertex for not too low temperature $T=0.2t$ and $n=1$. In this
case the chosen frequency box is sufficiently large (the maximal fermionic
frequency $\nu_{\mathrm{max}}\sim75t$) and the results calculated with and
without account of finite frequency box effects (we put $X_{q}^{s}=Z_{\nu
q}^{s}-1=\overline{\Phi}_{\nu\nu^{\prime}\omega}^{s}=0$ in the latter case) are close to each
other, with slightly better agreement of the result calculated with account of finite frequency box with the required asymptotic value.
With decreasing temperature to $T=0.08t$ the maximal fermionic frequency
$\nu_{\mathrm{max}}\sim30t$ and we observe stronger difference of the
fermion-boson vertex calculated with and without account of finite frequency box
effect (right part of Fig. \ref{FigGamma}; in this case we also change filling to $n=0.96$). The vertex,
evaluated with account of finite frequency box effects approaches correct
limiting value (equal to one). In both cases we find that the terms related to
the $\overline{\Phi}_{\nu\nu^{\prime}\omega}^{s}$ (i.e. second term in $X_{q}^{s}$ and the difference 
$Z_{\nu q}^{s}-1$) provide very small contribution ($<2\cdot 10^{-6}$ for $T=0.2t$ and $<10^{-4}$ for $T=0.08t$). We have also verified that the
obtained vertices $\gamma_{\nu q}^s$ yield the irreducible local susceptibility
$\phi_{\omega}^s$, obtained by the Eq. (\ref{phi1}) and the respective local susceptibility $\chi^s_\omega$, obtained by the Eq. (\ref{locvsir}), which agree with those
obtained directly from CT-QMC solver (for static local susceptibility at $T=0.2t$ we find $\chi^s_0=2.2074$ vs.~QMC result $2.2071$, while for $T=0.08t$ we find $\chi^s_0=3.77$ vs.~$3.78$, respectively). 

\begin{figure}[h!]
\center \includegraphics[width=0.7 \linewidth]{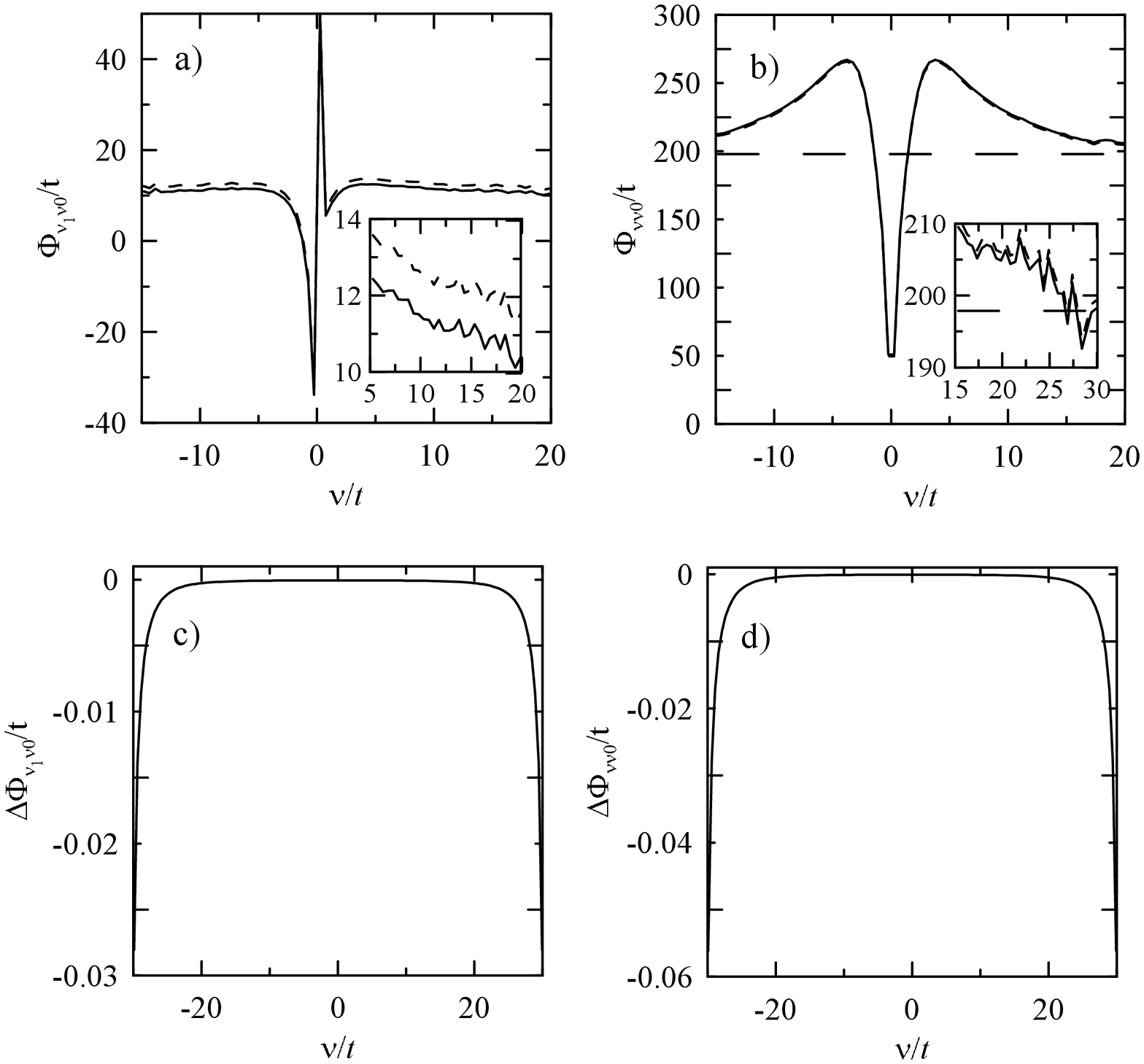} \endcenter
\caption{The 2PI vertex $\Phi^{s}_{\nu_{1}\nu0}$ (a, $\nu_{1}=\pi T$) and
$\Phi^{s}_{\nu\nu0}$ (b) in DMFT for two-dimensional Hubbard model
($t^{\prime}=0.15t$, $U=10t$) for $T=0.08t$ $n=0.96$. Short dashed line
corresponds to the calculation, performed only within the chosen frequency
box, while solid lines show the result according to the Eq. (\ref{Phi}). The
long dashed line in (b) shows limiting value $U+\overline{\Phi}_{\nu\nu0}^s$,
expected according to the Eq. (\ref{Vertex_ir_as}); insets show zoom of the asymptotic behavior. Plots (c) and (d) show contribution $\Delta \Phi_{\nu \nu' q}=\widetilde U_q^s (1-Z_{\nu q}^{s} Z_{\nu' q}^{s})$ to the third term in the right-hand side of Eq. (\ref{Phi}).}%
\label{FigPhi}%
\end{figure}

In Figs. \ref{FigPhi}a,b we show the frequency dependence of 2PI vertex $\Phi
_{\nu^{\prime}\nu0}^{s}$ at fixed frequency $\nu'=\nu_{1}=\pi T$ (left
part) and two equal frequencies $\nu=\nu^{\prime}$ (right part). For $\nu'=\nu_{1}$ one can
see that the obtained correction improves the high-frequency behavior, which
is close to $U$ in that case (the contribution $\overline{\Phi}_{\nu
\nu^{\prime}0}^s$ is small). At the same time, for $\nu=\nu
^{\prime}$ the obtained correction due to finite frequency box effect is
sufficiently small, and both vertices, with and without account of finite
frequency box effect approach the expected asymptotic value. The effect of the difference of the factors $Z_{\nu q}^{s}$ from unity in the third term in the right-hand side of Eq. (\ref{Phi}) remains sufficiently small, as can be seen from Figs. \ref{FigPhi} (c,d). The magnitude of the second vs. first term in $X_q^{s}$ is analyzed numerically below.  

\begin{figure}[h!]
\center \includegraphics[width=0.7 \linewidth]{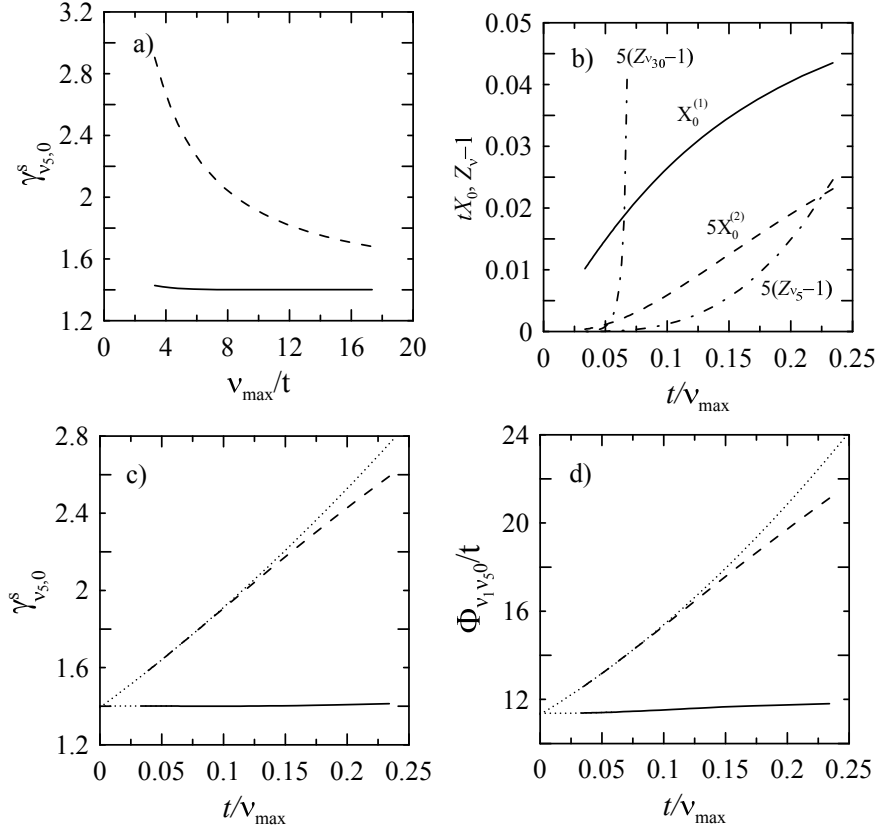} \endcenter
\caption{The dependence of (a,c) triangular vertex $\gamma^s_{\nu_{5},0}$, (b) first (solid line) and second (multiplied by 5, dashed line) term in $X_0^{s}$, as well as $Z_{\nu,0}^{s}-1$ (multiplied by 5) with $\nu=\nu_5$ and $\nu=\nu_{30}$ (dot-dashed lines), and (d) the 2PI vertex $\Phi^{s}_{\nu_{1}\nu 0}$ on $\nu_{\rm max}$ in DMFT for two-dimensional Hubbard model ($t^{\prime}=0.15t$, $U=10t$, $T=0.08t$, $n=0.96$, $\nu_{n}=(2n-1) \pi T$). Dashed lines in (a,c,d) correspond to the calculation, performed only within the chosen frequency box, while solid lines show the results according to the Eqs. (\ref{gamma}) and (\ref{Phi}). 
Dotted lines show extrapolation by quadratic polynomial (with respect to $1/\nu_{\rm max}$) of the results calculated at sufficiently large $\nu_{\rm max}$ without account of finite frequency box effects, and by $a+b/\nu_{\rm max}^4+c/\nu_{\rm max}^5$ of the results with account of frequency box effects.}%
\label{FigAs}%
\end{figure}

Finally, we verify the scaling of the obtained results with changing the size of the frequency box $\nu_{\rm max}$. For that we change the number of fermionic frequencies inside the frequency box between $18$ and $120$ and evaluate the respective vertices at fixed frequency $\nu$. The result of the calculation of dependence of both terms of $X_0^{s}$, the factors $Z_{\nu_{5},0}^s$ and $Z_{\nu_{30},0}^s$, triangular vertex $\gamma^s_{\nu_{5},0}$ and the two-particle irreducible vertex $\Phi^{s}_{\nu_{1}\nu_{5}0}$ ($\nu_n=(2n-1)\pi T$) on the size of the frequency box is shown in Fig. \ref{FigAs}. One can see that, as discussed in the end of previous Section, the first term in $X_0^{s}$ scales as $1/\nu_{\rm max}$ at large $\nu_{\rm max}$. At the same time, the second term in $X_0^{s}$ scales as $1/\nu_{\rm max}^2$, and, therefore, becomes negligibly small at sufficiently large $\nu_{\rm max}$. Although $Z_{\nu,0}^s-1\propto 1/\nu_{\rm max}^3$ decays faster than the second term in $X_0^s$, at intermediate $\nu_{\rm max}$ and $\nu\sim\nu_{\rm max}$ it becomes somewhat enhanced (cf. also Figs. \ref{FigPhi}c,d). As also discussed in the end of previous Section, the deviation of the vertices, calculated without account of finite frequency box effects from their values extrapolated  to $\nu_{\rm max}\rightarrow \infty$ (obtained using quadratic polynomial with respect to $1/\nu_{\rm max}$) scales as $1/\nu_{\rm max}$. At the same time, the vertices, calculated using the obtained formulae, change very weakly with $1/\nu_{\rm max}$ (we have verified that this holds for all $|\nu|<\nu_{\rm max}$). Using $a+b/\nu_{\rm max}^4+c/\nu_{\rm max}^5$ fits for vertices obtained with account of finite frequency box effects, we find the results of extrapolation consistent with those for vertices, obtained without of account of finite frequency box effects, which shows applicability of the obtained formulae. From these results it follows that for practical calculations without account of finite frequency box effects because of strong dependence of the vertices on $1/\nu_{\rm max}$, at least three  different sufficiently large sizes of frequency box should be considered to determine the coefficients of the quadratic polynomial, and, therefore, extrapolated values of the vertices. At the same time, since the results obtained with account of finite frequency box effects change very weakly with frequency box size, only one such calculation is sufficient with reasonable accuracy in that case. 

\section{Conclusion}

In conclusion, we have derived explicit formulae for the full (Eq.
(\ref{Lambda_box})) and reduced (Eqs. (\ref{gamma}), (\ref{gamma1})) fermion-boson vertices; full
(Eq. (\ref{chi1})) and irreducible (Eq. (\ref{phi})) susceptibilities, and the
2PI vertex (\ref{Phi}), which contain summation only in a given frequency box.
These formulae account for the contribution of the frequencies outside the
frequency box via the terms, containing $X_{q}^{c(s)}$ and $Z_{\nu q}^{c(s)}$. In contrast to the approach, which neglects the corrections due to finiteness of the frequency box, which results error scales as $1/\nu_{\rm max}$, the considered approach is expected to show  $1/\nu_{\rm max}^4$ scaling of the error, and requires therefore rather small sizes of the frequency box.
We have verified numerically applicability of the obtained results
on the two-dimensional Hubbard model with next-nearest hopping and strong
Coulomb repulsion.

The obtained results can be used in a broad range of applications of
diagrammatic extensions of dynamical mean field theory.

\textit{Acknowledgements. } The work is partly supported by the theme ``Quant"
AAAA-A18-118020190095-4 of Minobrnauki, Russian Federation. The calculations are performed on the ``Uran" cluster of UB RAS.

\end{document}